\documentstyle[aps,eqsecnum,preprint,epsf]{revtex}
\begin{document} 
\preprint{cond-mat/9706074}
\draft
\title{Measurements with a noninvasive detector and 
dephasing mechanism}
\author{S.A. Gurvitz}

\address{Department of Particle Physics, Weizmann Institute of
         Science, Rehovot 76100, Israel}
\date{\today}
\maketitle
\begin{abstract}
We study dynamics of the measurement process in 
quantum dot systems, where a particular state out of 
coherent superposition is observed. The ballistic point-contact 
placed near one of the dots is taken as a noninvasive detector. 
We demonstrate that the measurement process is fully described by 
the Bloch-type equations applied to the whole system. 
These equations clearly reproduce the collapse of the 
density-matrix into the statistical mixture in the course of 
the measurement process. The corresponding dephasing 
width is uniquely defined.  We show that 
the continuous observation of one of the states 
in a coherent superposition may {\em accelerate} decay from this state -- 
in contradiction with rapidly repeated observations, which slow down  
the transitions between quantum states 
(the quantum Zeno effect). 
\end{abstract}
\section{introduction}
In recent years there have been many measurements in mesoscopic 
systems sensitive to the phase of electronic wave function.
We mention the experiments with double split systems\cite{ya1,buks1}, 
quantum dot embedded in
Aharonov-Bohm ring\cite{ya2,buks2}, and coupled quantum dots\cite{vdr}.
It is known that the phase of wave function, or more precisely the 
off-diagonal density matrix elements, can be destroyed by interaction 
with the environment or with the measurement device. As a result, 
the density matrix becomes the statistical mixture. The 
latter does not display any coherence effects.
Now the rapid progress in microfabrication technology allows 
us to investigate experimentally the dephasing process in mesoscopic 
systems, for instance by observation of 
a particular state out of coherent superposition\cite{buks3}.

Although the dephasing (decoherence) plays important role   
in different processes, its mechanism is not elaborated enough. 
For instance, in many studies of the quantum measurement problems 
the dephasing is usually accounted for by introducing some 
phenomenological dissipating terms, associated 
with a detector (or an environment). Yet, such a procedure 
cannot not illuminate the origin of the dephasing and its role 
in the measurement problem. Most appropriate way to approach the 
problem, however, is to start with the microscopic description 
of the measured system and the detector together
by use of the Schr\"odinger equation, 
$i\dot\sigma =[{\cal H},\sigma ]$, where  
$\sigma ({\cal S},{\cal S}';{\cal D},{\cal D}',t)$ is the 
total density-matrix and ${\cal H}$ is the Hamiltonian 
for the entire system. Here ${\cal S}({\cal S}')$ and 
${\cal D}({\cal D}')$ are the variables 
of the measured system and the detector respectively. In this case the 
influence of the detector on the measured system can be determined 
by ``tracing out'' the detector variables in the total density matrix,
\begin{equation}
\sum_D\sigma ({\cal S},{\cal S}',{\cal D},{\cal D},t)\to
\bar\sigma ({\cal S},{\cal S}',t).
\label{in1}
\end{equation}
The decoherence would correspond to an exponential damping 
of the off-diagonal 
matrix elements in the reduced density-matrix: 
$\bar\sigma ({\cal S},{\cal S}',t)\sim 
\exp (-\Gamma_dt)\to 0$ for ${\cal S}\not ={\cal S}'$,
with $\Gamma_d$ is the decoherence rate. 

In this paper we apply the above approach to study the decoherence, generated 
by measurement of a quantum-dot occupancy in multi-dot systems.
As the measurement device (detector) 
we take the ballistic point-contact in 
close proximity to the measured quantum dot\cite{pep}.
Since the quantum-mechanical description of this 
detector is rather simple, it allows us to investigate
the essential physics of the measurement process in great details.  
In addition, the ballistic point-contact 
is a noninvasive detector\cite{pep}. 
Indeed, the time which an electron 
spends inside it is very short. Thus, the point-contact   
would not distort the measured dot. (The first measurement of 
decoherence in the quantum dot generated by the point-contact 
has been recently performed by Buks {\em et al.}\cite{buks3}). 

The plan of this paper is the following: In Sect. 2 we describe the 
measurement of a quantum-dot occupation, when the current 
flows through this dot. We use the quantum rate 
equations\cite{gp,g,glaz,ak,been}, 
which allow us to describe both, 
the measured quantum dot and the point-contact detector in the most simple
way. Detailed microscopic derivation 
of the rate equations for the point-contact 
is presented in Appendix A. In Sect. 3 we investigate the decoherence 
of an electron in a double-well potential caused by the point-contact 
detector by measuring  the occupation of one of the wells. 
Special attention is paid to comparison with the result of rapidly 
repeated measurement. For a description of this system we use 
the Bloch-type rate equations\cite{gp,g,naz},  which are derived in Appendix B. 
Similar decoherence effects, but in dc current flowing through 
a coupled-dot system are discussed in Sect. 4. The last section is summary.

\section{Ballistic point-contact detector}
Consider the measurement of electron occupation 
of a semiconductor quantum dot by
means of a separate measuring circuit in close proximity\cite{buks3,pep}. 
A ballistic one-dimensional point-contact is used as a ``detector'' that
resistance is very sensitive to the electrostatic field generated 
by an electron occupying the measured quantum dot. Such a set up is shown 
schematically in Fig.~1, where the detector is represented by  
a barrier, connected with two reservoirs at the chemical potentials 
$\mu_L$ and $\mu_R$ respectively. The transmission probability 
of the barrier varies from $T$ to $T'$, depending on whether or not 
the quantum dot is occupied by an electron, Fig.~1 (a,b).

Initially all the levels in the reservoirs are filled up to the  
corresponding Fermi energies and the quantum dot is empty. 
(For simplicity we consider the reservoirs at zero temperature).
Such a state is not stable, since electrons are moving from the left to the 
right reservoir. The time-evolution of the entire system can 
be described by the master (rate) equations\cite{gp,g,glaz,ak,been} 
(the microscopic derivation from the many-body Schr\"odinger equation 
is given in Appendix A and in Refs.\cite{gp,g}). 

\vskip1cm
\begin{minipage}{13cm}
\begin{center}
\leavevmode
\epsfxsize=13cm
\epsffile{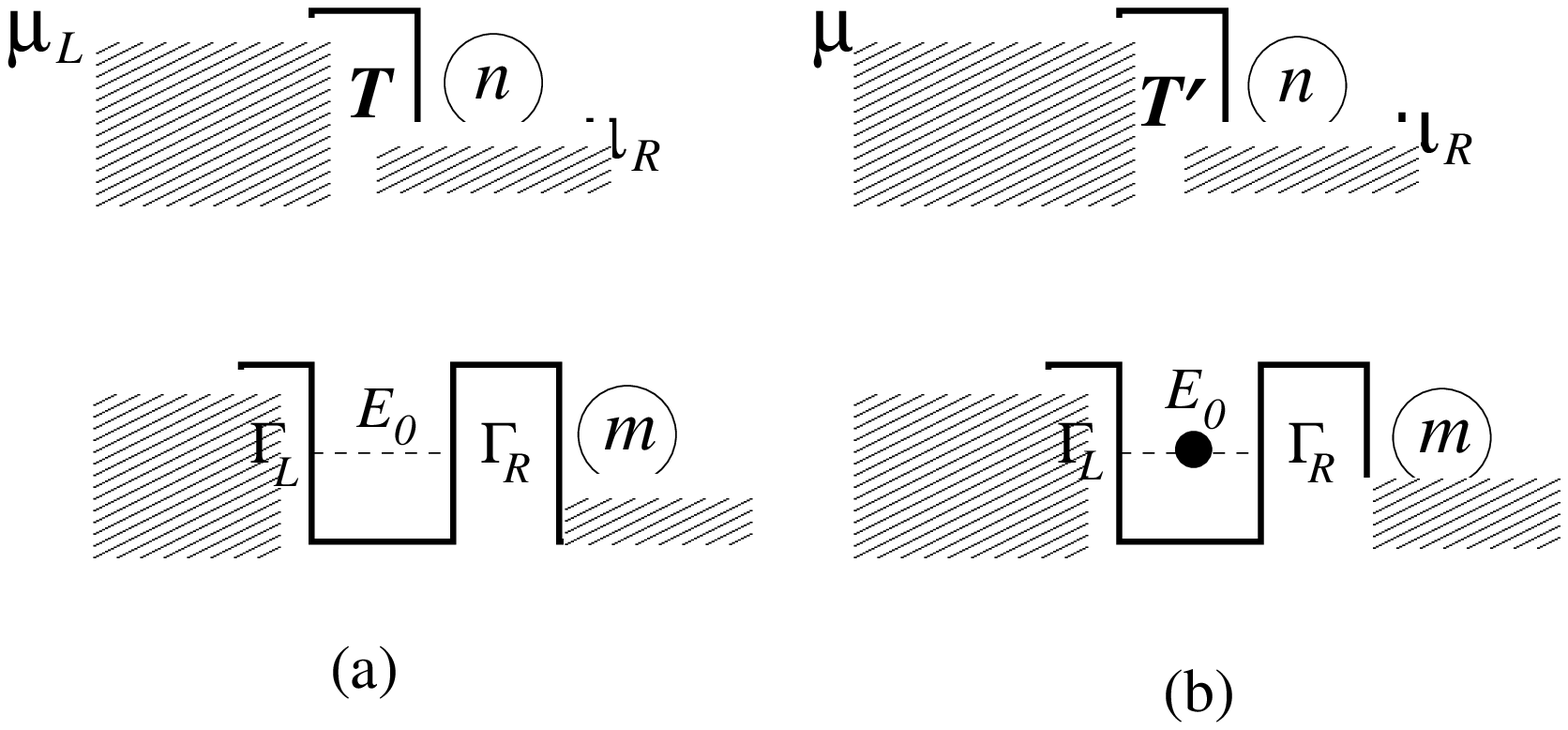}
\end{center}
{\begin{small}
Fig.~1. Ballistic point-contact near quantum-dot. $\Gamma_{L,R}$ 
are the corresponding tunneling rates. The penetration coefficient
of the point-contact is $T$ for the empty dot (a) and $T'$ for the occupied 
dot (b). The indices $m$ and $n$ denote the number of electrons 
penetrating to the right reservoirs at time $t$.   
\end{small}}
\end{minipage} \\ \\ 
In order to write down these equations we introduce the 
probabilities $\sigma_{aa}^{m,n}(t)$ and $\sigma_{bb}^{m,n}(t)$  
of finding the entire system in the states $|a\rangle$ and 
$|b\rangle$ corresponding to empty or occupied dot (Fig.~1 a,b).
Here $m$ and $n$ are the number of electrons penetrated 
to the right reservoirs of the measured system and the detector, 
respectively. The corresponding rate equations for these 
probabilities have the following form
\begin{mathletters}
\label{a1}
\begin{eqnarray}
\dot\sigma_{aa}^{m,n}&=&-(\Gamma_L+D)\sigma_{aa}^{m,n}
+\Gamma_R\sigma_{bb}^{m-1,n}+D\sigma_{aa}^{m,n-1}
\label{a1a}\\
\dot\sigma_{bb}^{m,n}&=&-(\Gamma_R+D')\sigma_{bb}^{m,n}
+\Gamma_L\sigma_{aa}^{m,n}+D'\sigma_{bb}^{m,n-1},
\label{a1b}
\end{eqnarray}
\end{mathletters}
where $\Gamma_{L,R}$ are the transition rates for an
electron tunneling from the left reservoir to the dot 
and from the dot to the right reservoir respectively,
and $D=T(\mu_L-\mu_R)/2\pi$ is the rate of electron hopping 
from the right to the left reservoir through the point-contact  
(the Landauer formula). 

The accumulated charge in the right reservoirs of the detector ($d$)
and of the measured system ($s$) is given by  
\begin{mathletters}
\label{a2}
\begin{eqnarray}
Q_d(t)&=&\sum_{m,n}n[\sigma_{aa}^{m,n}(t)+\sigma_{bb}^{m,n}(t)]
\label{a2a}\\
Q_s(t)&=&\sum_{m,n}m[\sigma_{aa}^{m,n}(t)+\sigma_{bb}^{m,n}(t)]
\label{a2b}
\end{eqnarray}
\end{mathletters}
(We choose the units where the electron charge $e=1$, and $\hbar=1$).
The currents flowing in the detector and in the measured system
are $I_d(t)=\dot Q_d(t)$ and $I_s(t)=\dot Q_s(t)$. 
Using Eqs.~(\ref{a1}) and (\ref{a2}) we obtain
\begin{mathletters}
\label{aa3} 
\begin{eqnarray}
I_d(t) &=& \sum_{m,n}n[\dot\sigma_{aa}^{m,n}(t)
+\dot\sigma_{bb}^{m,n}(t)]=
D\sigma_{aa}(t)+D'\sigma_{bb}(t),
\label{a5}\\
I_s(t) &=& \sum_{m,n}m[\dot\sigma_{aa}^{m,n}(t)
+\dot\sigma_{bb}^{m,n}(t)]=
\Gamma_R\sigma_{bb}(t)\ ,
\label{a3}
\end{eqnarray}
\end{mathletters}
where
$\sigma_{aa}\equiv\sum_{m,n}\sigma_{aa}^{m,n}$ and 
$\sigma_{bb}\equiv\sum_{m,n}\sigma_{bb}^{m,n}$ are the total 
probabilities of finding the dot empty or occupied. Obviously 
$\sigma_{aa}(t)=1-\bar\sigma (t)$, where $\bar\sigma (t) 
\equiv \sigma_{bb} (t)$.   
Performing the summation over $m,n$ in Eqs.~(\ref{a1}) we obtain the 
following rate equation for the quantum dot occupation probability 
$\bar\sigma$
\begin{equation}
\dot{\bar\sigma}(t)=\Gamma_L -(\Gamma_L+\Gamma_R)\bar\sigma (t)\ .
\label{a4}
\end{equation}

If the point-contact and the quantum dot are decoupled, 
the detector current is $I_d^{(0)}=D$.
Hence, the occupation of the quantum dot can be measured 
through the variation of the detector current $\Delta I_d = I_d^{(0)}-I_d$. 
One readily obtains from Eq.~(\ref{a5}) that 
\begin{equation}
\Delta I_d(t)=\frac{\Delta T\, V_d}{2\pi}\bar\sigma (t),
\label{a8}
\end{equation}
where $V_d=\mu_L-\mu_R$ is the voltage bias, and $\Delta T=
T-T'$. Thus, the point contact is indeed the measurement device.
In fact, Eq.~(\ref{a8}) is a self-evident one. Indeed, the variation 
of the point-contact current is $\Delta TV_d/2\pi$ and $\bar\sigma$ 
is the probability for such a variation.  
 
It follows from Eqs.~(\ref{a1}), (\ref{aa3}) that the same current  
$I_s(t)=\Gamma_R\bar\sigma (t)$ would flow through the quantum dot 
in the absence of the detector ($D=D'=0$). 
It means that the point-contact detector is a noninvasive detector.
This is not surprising since only an electron inside the point-contact
(under the barrier) can affect an electron in the quantum dot.
The relevant (tunneling) time is very short. Actually, it 
is zero in the tunneling Hamiltonian approximation, 
Eqs.~(\ref{ap1}), (\ref{bp1}), used for the derivation 
of the rate equations. 

\section{Detection of electron oscillations in coupled-dots}
A well-known manifestation of quantum coherence is   
the oscillation of a particle in a double-well (double-dot) potential. 
The origin of these oscillations is the interference between the
probability amplitudes of finding a particle in different wells.
Hence, one can expect that the disclosure of a particle  
(electron) in one of wells would generate the ``dephasing'' 
that eventually destroys these oscillations,
even without distorting the energy levels of the system. 

Let us investigate the mechanism of this process by taking for detector   
a noninvasive point-contact. A possible set up is shown in Fig. 2.   
\vskip1cm
\begin{minipage}{13cm}
\begin{center}
\leavevmode
\epsfxsize=13cm
\epsffile{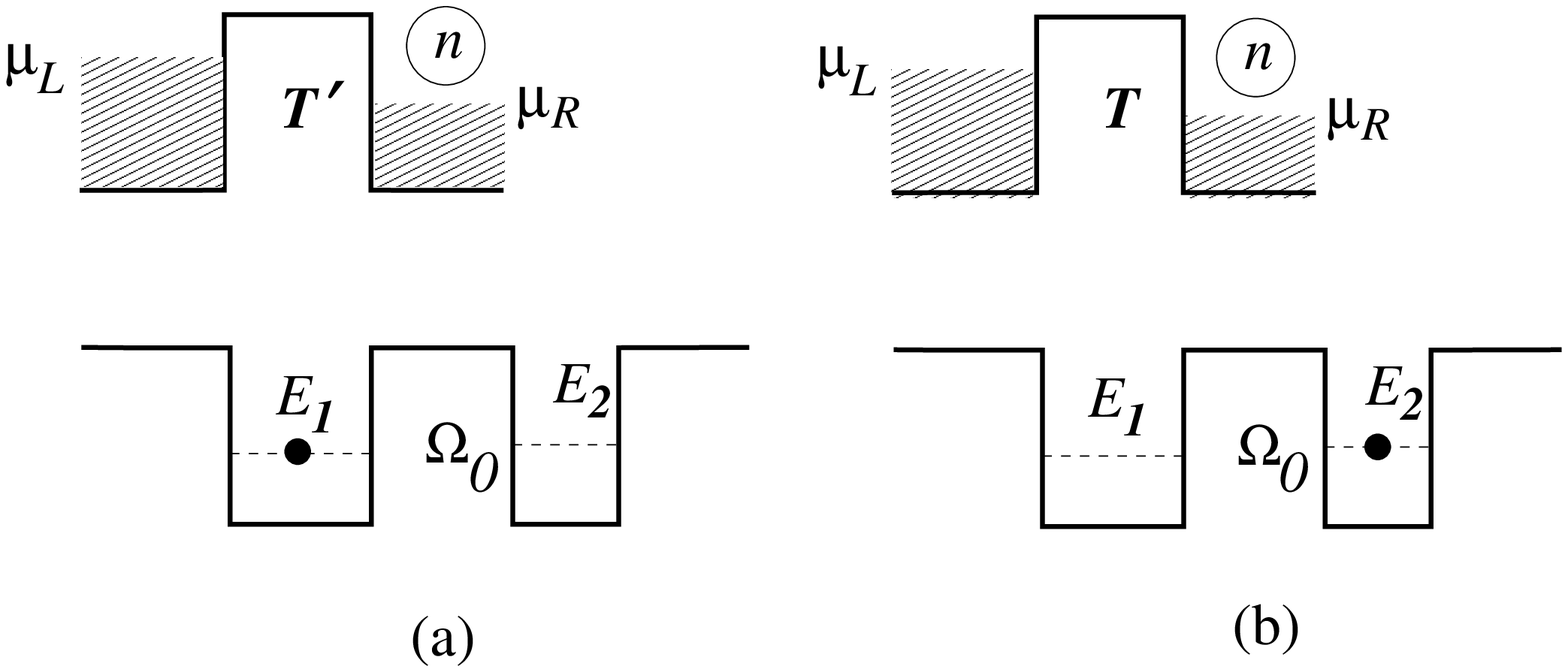}
\end{center}
{\begin{small}
Fig.~2. Electron oscillations in the double-well. The penetration coefficient
of the point-contact varies from $T'$ to $T$ when an electron occupies the
left well (a) or the right well (b), respectively. The index $n$ denotes 
the number of electrons accumulated in the collector at time $t$. 
\end{small}}
\end{minipage} \\ \\ 
We assume that the transmission probability of the point-contact is $T$ 
when an electron occupies the right well, and it is $T'$ when
an electron occupies the left well. Here $T'<T$ since   
the right well is away from the point contact.

Now we apply the quantum-rate equations\cite{gp,g} to the whole system.
However, in the distinction with the previous case, the electron 
transitions in the measured system take place  
between the {\em isolated} states inside the dots.
As a result the diagonal density-matrix elements are coupled with the
off-diagonal elements, so that the corresponding rate equations are
the Bloch-type equations\cite{gp,g,naz}. 

We first start with the case of the double-well detached 
from the point-contact detector. 
The Bloch equations describing the time evolution of the electron 
density-matrix $\sigma_{ij}$ have the following form 
\begin{mathletters}
\label{c1}
\begin{eqnarray}
\dot\sigma_{aa}& = &i\Omega_0 (\sigma_{ab}-\sigma_{ba})\;, 
\label{c1a}\\
\dot\sigma_{bb}& = &i\Omega_0 (\sigma_{ba}-\sigma_{ab})\;,
\label{c1b}\\
\dot\sigma_{ab}& = & i\epsilon\sigma_{ab}+i\Omega_0(\sigma_{aa}
-\sigma_{bb}),
\label{c1c}
\end{eqnarray}
\end{mathletters}
where $\epsilon =E_2-E_1$ and $\Omega_0$ 
is the coupling between the left and the right wells. 
Here $\sigma_{aa}(t)$ and $\sigma_{bb}(t)$ are 
the probabilities of finding the electron in the left and the 
right well respectively, and $\sigma_{ab}(t)=\sigma_{ba}^*(t)$
are the off-diagonal density-matrix elements (``coherences'')\cite{bloch}.

Solving these equations for the initial conditions and $\sigma_{aa}(0)=1$ and
$\sigma_{bb}(0)=\sigma_{ab}(0)=0$ we obtain 
\begin{equation}
\sigma_{aa}(t)=\frac{\Omega_0^2\cos^2(\omega t)+\epsilon^2/4}{\Omega_0^2+
\epsilon^2/4},
\label{c2}
\end{equation}
where $\omega =(\Omega_0^2+\epsilon^2/4)^{1/2}$. As expected the electron
initially localized in the first well oscillates between the wells
with the frequency $\omega$. Notice that the amplitude of these 
oscillations is $\Omega_0^2/(\Omega_0^2+\epsilon^2/4)$.
Thus the electron remains localized in the first well if  
the level displacement is large, $\epsilon \gg\Omega_0$. 

Now we consider the electron oscillations in the presence of  
the point contact detector, Fig. 2. 
The corresponding Bloch equations for the entire system have the 
following form (Appendix B):
\begin{mathletters}
\label{c3}
\begin{eqnarray}
\dot\sigma_{aa}^{n} & = & -D'\sigma_{aa}^{n}+D'\sigma_{aa}^{n-1}
+i\Omega_0 (\sigma_{ab}^{n}-\sigma_{ba}^{n})\;,
\label{c3a}\\
\dot\sigma_{bb}^{n} & = & -D\sigma_{bb}^{n}
+D\sigma_{bb}^{n-1}-i\Omega_0 (\sigma_{ab}^{n}-\sigma_{ba}^{n})\;,
\label{c3b}\\
\dot\sigma_{ab}^{n} & = & i\epsilon\sigma_{ab}^{n}+
i\Omega_0(\sigma_{aa}^{n}-\sigma_{bb}^{n})
-\frac{1}{2}(D'+D)\sigma_{ab}^{n}
+(D\, D')^{1/2}\sigma_{ab}^{n-1}\, ,
\label{c3c}
\end{eqnarray}
\end{mathletters} 
Here the index $n$ denotes the number of electrons arriving to the 
collector at time $t$, and $D(D')$ is the transition rate of an electron 
hopping from the left to the right detector reservoirs,  
$D=T(\mu_L-\mu_R)/2\pi$, Eqs.~(\ref{a1}). 
Notice that the presence of the detector
results in additional terms in the rate equations in comparison with
Eqs.~(\ref{c1}). These terms are generated by transitions of 
an electron from the left to the right detector reservoirs with 
the rates $D$ and $D'$ respectively. 
The equation for the non-diagonal 
density-matrix elements $\sigma_{ab}^{n}$
is slightly different from the standard Bloch equations
due to the last term, which describes transition between different 
coherences, $\sigma^{n-1}_{ab}$ and $\sigma^{n}_{ab}$. 
This term appears in the Bloch equations for coherences 
whenever the same hopping ($n-1\to n$) takes place in the {\em both} 
states of the off-diagonal density-matrix element ($a$ and $b$)
(see Refs.\cite{gp,g} and Appendix B).
The rate of such transitions is determined by 
a product of the corresponding {\em amplitudes} ($T^{1/2}$ and 
${T'}^{1/2}$).

It follows from Eqs.~(\ref{a5}), (\ref{c3}) 
that the variation of the point-contact  
current $\Delta I_d(t)=
I^{(0)}-I_d(t)$ measures directly the charge in 
the first dot. Indeed, one obtains for the detector current
\begin{equation}
I_d(t)=\sum_nn[\sigma_{aa}^n(t)+\sigma_{bb}^n(t)]=D'\sigma_{aa}(t)+
D\sigma_{bb}(t),
\label{cc3}
\end{equation}
where $\sigma_{ij}=\sum_n\sigma_{ij}^n$. Therefore $\Delta I_d(t)$ is given 
by Eq.~(\ref{a8}), where $\bar\sigma (t)\equiv \sigma_{aa}(t)$.

In order to determine the influence of the detector on the double-well 
system we trace out the detector states in Eqs.~(\ref{c3})
thus obtaining
\begin{mathletters}
\label{c4}
\begin{eqnarray}
\dot\sigma_{aa}& = &i\Omega_0(\sigma_{ab}-\sigma_{ba})\;, 
\label{c4a}\\
\dot\sigma_{bb}& = &i\Omega_0(\sigma_{ba}-\sigma_{ab})\;,
\label{c4b}\\
\dot\sigma_{ab}& = & i\epsilon\sigma_{ab}+i\Omega_0(\sigma_{aa}
-\sigma_{bb})-\frac{1}{2}(\sqrt{D}-\sqrt{D'})^2\sigma_{ab},
\label{c4c}
\end{eqnarray}
\end{mathletters}
where $\sigma_{ij}=\sum_n\sigma^{n}_{ij}(t)$. 

These equations coincide with   
Eqs.~(\ref{c1}), describing the electron oscillations without 
detector, except for the last term in Eq.~(\ref{c4c}). The latter 
generates the exponential damping of the non-diagonal density-matrix 
element with the ``dephasing'' rate\cite{fn1} 
\begin{equation}
\Gamma_d=(\sqrt{D}-\sqrt{D'})^2=(\sqrt{T}-\sqrt{T'})^2\frac{V_d}{2\pi}
\label{c5}
\end{equation}
It implies that $\sigma_{ab}\to 0$ for $t\to\infty$. We can check it 
by looking for the stationary solutions of Eqs.~(\ref{c4}) in 
the limit $t\to\infty$. In this case $\dot\sigma_{ij}(t \to\infty )=0$
and Eqs.~(\ref{c4}) become 
linear algebraic equations, which can be easily solved.   
One finds that the electron density-matrix becomes 
the statistical mixture. 
\begin{equation} 
\sigma (t)=\left (\begin{array}{cc}
\sigma_{aa}(t)&\sigma_{ab}(t)\\
\sigma_{ba}(t)&\sigma_{bb}(t)\end{array}\right )
\to\left (\begin{array}{cc}
1/2&0\\0&1/2
\end{array}\right ) \;\; {\mbox{for}}\;\;\;  t\to\infty.
\label{c6}
\end{equation}
Notice that the damping of the nondiagonal density matrix elements 
is coming entirely from the possibility of disclosing the electron 
in one of the wells. Indeed, if the detector does not 
distinguish which of the wells is occupied, 
i.e. $T=T'$, then $\Gamma_d=0$. 

The Bloch equations (\ref{c3}), (\ref{c4}) display explicitly the
mechanism of the dephasing during a noninvasive measurement,
i.e. that which does not distort 
the energy levels of the measured system\cite{fn2}.
The dephasing appears in the reduced density matrix as the ``dissipative'' 
term in the nondiagonal density matrix elements only,
as a result of tracing out the detector variables. All other terms 
related to the detector are canceled after tracing out the detector 
variables. It is important to note that  such a dephasing term 
in Eq.~(\ref{c4c}) generates the ``collapse'' of the electron density 
matrix into the statistical mixture, Eq.~(\ref{c6}), without explicit 
use of the measurement reduction postulate\cite{neu}. 
The collapse is fully described by the Bloch-type equations, derived 
from the Shr\"odinger equation (Appendix B).

In fact, the idea that the dissipative interaction of a measured system 
with a detector can be responsible for the density matrix collapse 
is not new. It was discussed in many publications, as
for instance in works of Zurek\cite{zur}, which stressed 
conceptual points, or in detailed studies of more specific 
examples of atomic transitions\cite{beige}. Yet, the present 
study of mesoscopic systems elaborates additional aspects of 
the dephasing problem. These are the dephasing 
mechanism due to continuous observation
with a non-invasive detector, and the striking difference between 
the continuous and rapidly repeated measurements. 
The latter is discussed below.

\subsection{Continuous measurement and Zeno effect}
The most surprising phenomenon which displays Eq.~(\ref{c6}) is that 
the transition to the statistical mixture
takes place even for a large displacement of the energy levels, 
$\epsilon\gg\Omega_0$, irrespectively of the initial conditions. 
It means that an electron initially localized in one of the
wells would be always {\em delocalized} at $t\to\infty$.  
It would happened even if the electron was initially localized 
at the lower level. (Of course it does not 
violate the energy conservation, since the double-well 
is not isolated). Such a behavior 
is not expectable because the amplitude of electron 
oscillations is very small for large level displacement,   
Eq.~(\ref{c2}). Thus, the electron should 
stay localized in one of the wells. 
One could expect that the continuous observation of this electron 
by a detector could only increase its localization. 
It can be inferred from so called Zeno effect\cite{zeno}.
The latter tells us that repeated observation of the system 
slow down transitions between quantum states due to the collapse 
of the wave function into the observed state. Since in our case 
the change of the detector current, $\Delta I_s(t)$  
monitors $\bar\sigma (t)$ in the left well,
Eq.~(\ref{a8}), (\ref{cc3}), it represents the continuous measurement 
of the charge in this well. Nevertheless the effects is just opposite --
the continuous measurement delocalizes the system\cite{fn3}. 

In fact, our results for small $t$ 
seems to be in an agreement with the Zeno effect, even so 
we have not explicitly implied the projection postulate. 
For instance,  Fig. 3a shows the time-dependence of the probability to 
find an electron in the left dot, as obtained from the 
solution of Eqs.~(\ref{c4}) for the aligned levels ($\epsilon =0$), and 
$\Gamma_d=0$ (dashed curve), $\Gamma_d=4\Omega_0$ (dot-dashed 
curve) and  $\Gamma_d=16\Omega_0$ (solid curve). One finds that 
for small $t$ the rate of transition from the left to the right well  
decreases with the increase of $\Gamma_d$. 
\vskip1cm
\begin{minipage}{13cm}
\begin{center}
\leavevmode
\epsfxsize=8cm
\epsffile{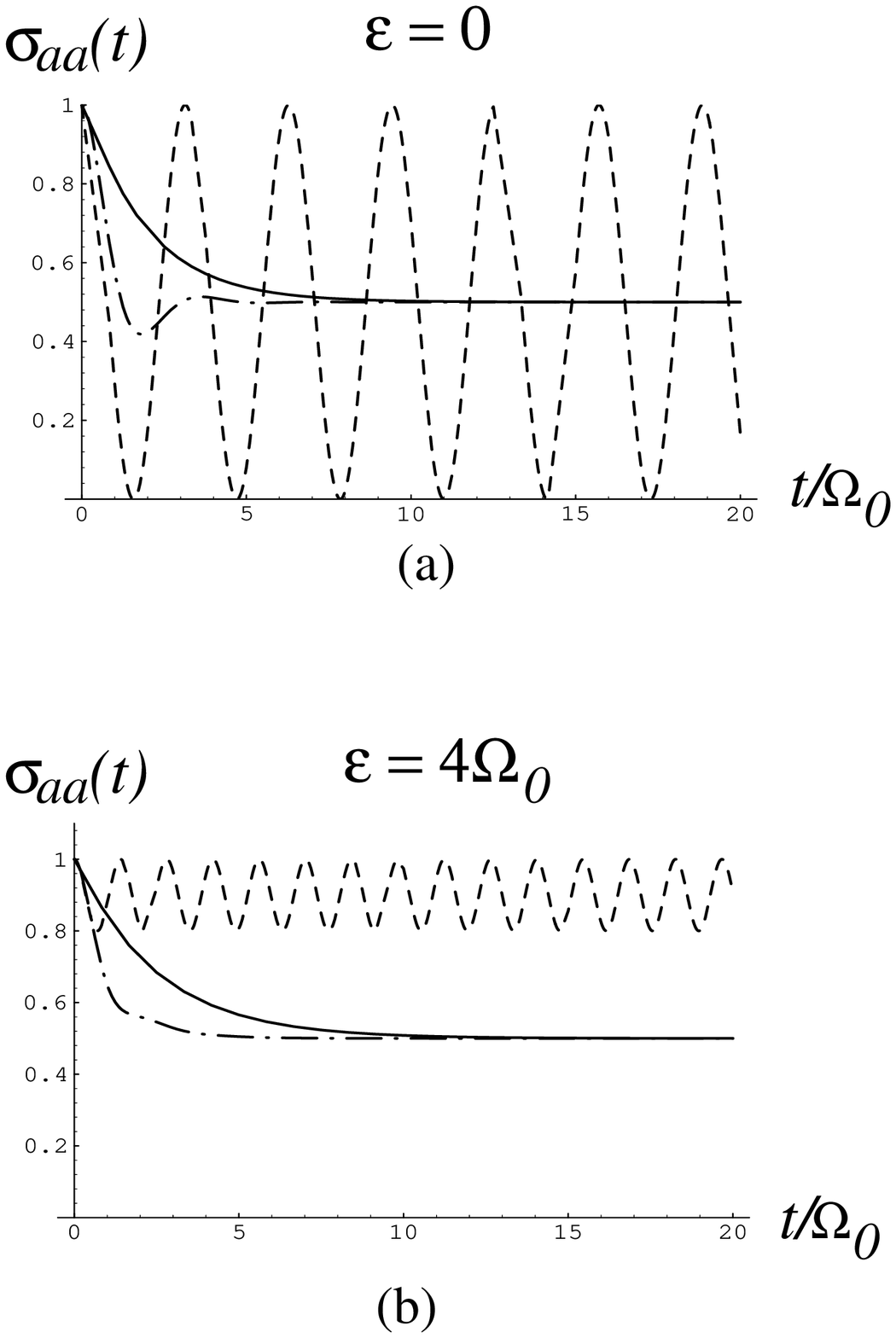}
\end{center}
{\begin{small}
Fig.~3. The occupation of the first well as a function 
of time, Eqs. (\ref{c4}): 
(a) the levels are aligned ($\epsilon =0$); (b) the 
levels are displaced ($\epsilon =4\Omega_0$). The curves 
correspond to different values of the dephasing rate:   
$\Gamma_d=0$ (dashed),  
$\Gamma_d=4\Omega_0$ (dot-dashed),  and $\Gamma_d=16\Omega_0$ 
(solid).
\end{small}}
\end{minipage} \\ \\ 

The same slowing down of the transition rate for small $t$
we find for the disaligned  levels ($\epsilon =4\Omega_0$) in Fig. 3b. 
It implies that very frequent repeated measurements  
would indeed localize the system. 
In that sense the Bloch equations reproduce 
the Zeno effect without explicit use of the projection postulate.
Actually, this result has been found earlier by an analysis of atomic 
transitions by using 
the Bloch equation for 3-level system\cite{fre,fn4}.
It was shown that the repeated measurement with short intervals 
$\Delta t =t/n$ localizes the system in the limit $n\to\infty$. 
Yet in our case the continuous measurement leads to an electron 
{\em delocalization}, whereas in 
the absence of detector an electron would stay localized 
in the left well (the dashed curve in Fig. 3b).
Thus the continuous and very frequent repeated measurements 
affect the system in opposite ways. 

Our microscopic treatment allows us to determine the origin of 
the difference in both treatments. One easily finds that 
the derivation of the Bloch-type of equations, describing 
the measured system, Eqs.~(\ref{c4}) implies the tracing of 
the detector variables, Eq.~(\ref{in1}). Since this procedure 
is outside the Schr\"odinger equation, it could distort the time 
development of the system. In our case of continuous 
measurement the tracing is done at the time $t$, whereas the   
frequent repeated measurement with the intervals $\Delta t= t/n$ implies
that the tracing of the detector variables takes part at the end of each 
interval $\Delta t$. As a result the limit of $n\to\infty$ the measured 
system stays localized\cite{fre}.  

\section{Measurement of resonant current in coupled-dots}
In spite of great progress made in the microfabrication technique,
the direct measurement of single electron oscillations in coupled-dot
system is still a complicated problem. However, it is 
much easier to measure similar quantum coherence 
effects in electrical current flowing 
through coupled-dot systems. We therefore consider the same coupled-dot
of the previous section,  
but now connected with two reservoirs (emitter and collector).
As in the previous 
example the point-contact detector measures the occupation of the 
left dot, Fig. 4. 
For the sake of simplicity we assume strong  
inner and inter-dot  Coulomb repulsion, so only one electron can occupy  
this system\cite{naz}. 
Then there are only three available states 
of the coupled-dot system: the dots are empty ($a$),  
the first dot is occupied ($b$) 
and the second dot is occupied ($c$).    
In an analogy with Eqs.~(\ref{a1}), (\ref{c3}) we write the 
following Bloch equations for the density matrix $\sigma^{m,n}_{ij}(t)$
describing the entire system\cite{gp,g}: 
\begin{mathletters}
\label{b5}
\begin{eqnarray}
\dot\sigma_{aa}^{m,n} & = & -(\Gamma_L+D)\sigma_{aa}^{m,n}
+\Gamma_R\sigma_{cc}^{m-1,n}+D\sigma_{aa}^{m,n-1}\;,
\label{b5a}\\
\dot\sigma_{bb}^{m,n} & = & -D'\sigma_{bb}^{m,n}+D'\sigma_{bb}^{m,n-1}
+\Gamma_L\sigma_{aa}^{m,n}
+i\Omega_0 (\sigma_{bc}^{m,n}-\sigma_{cb}^{m,n})\;,
\label{b5b}\\
\dot\sigma_{cc}^{m,n} & = & -(\Gamma_R+D)\sigma_{cc}^{m,n}
+D\sigma_{cc}^{m,n-1}-i\Omega_0 (\sigma_{bc}^{m,n}-\sigma_{cb}^{m,n})\;,
\label{b5c}\\
\dot\sigma_{bc}^{m,n} & = & i\epsilon\sigma_{bc}^{m,n}+
i\Omega_0(\sigma_{bb}^{m,n}-\sigma_{cc}^{m,n})
-\frac{1}{2}(\Gamma_R+D'+D)\sigma_{bc}^{m,n}
+(D\, D')^{1/2}\sigma_{bc}^{m,n-1}\, ,
\label{b5d}
\end{eqnarray}
\end{mathletters}
where the indices $n$ and $m$ denote the number of electrons arrived 
at time $t$ to the upper and the lower collector reservoir, respectively. 
Here  $\Gamma_{L}$, $\Gamma_{R}$ are the rates of electron transitions 
from the left reservoir to the first dot and from the second dot 
to the right reservoir, and $\Omega_0$ is the amplitude of hopping between 
two dots. 
\vskip1cm
\begin{minipage}{13cm}
\begin{center}
\leavevmode
\epsfxsize=8cm
\epsffile{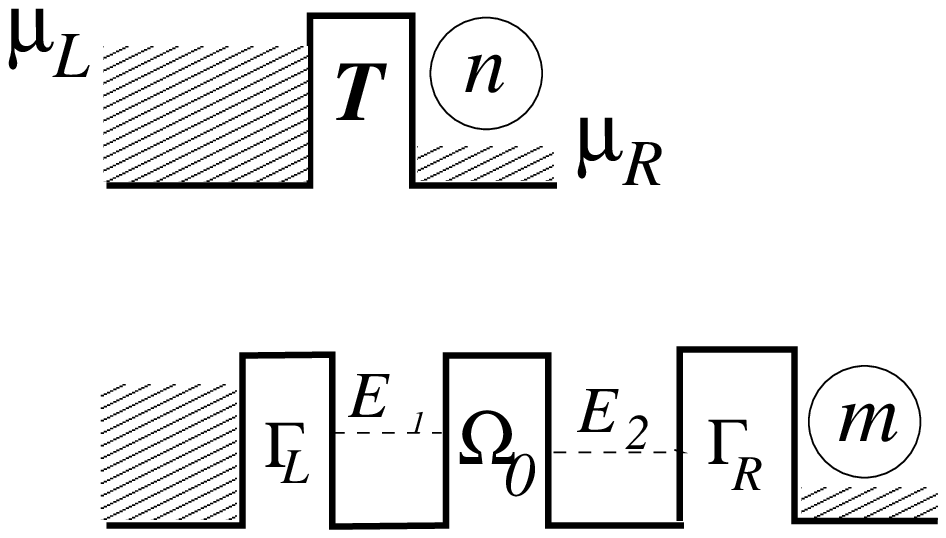}
\end{center}
{\begin{small}
Fig.~4. Resonant tunneling through the double-dot. $\Gamma_{L,R}$ denote the 
corresponding rate for the tunneling from (to) the left (right) reservoirs. 
The penetration coefficient of the point-contact is $T$ for the empty 
double-dot system or for the occupied second dot, 
and it is $T'$ for the occupied first dot. The indices $m$ and $n$ 
denote the number of electrons penetrating to the right reservoirs at 
time $t$.  
\end{small}}
\end{minipage} \\ \\ 

The currents in the double-dot system ($I_s$) and in the detector ($I_d$) 
are given by the following expressions (c.f. Eqs.~(\ref{a2}), (\ref{a3})):
\begin{mathletters}
\label{bb5}
\begin{eqnarray}
I_s&=&\sum_{m,n}m(\dot\sigma_{aa}^{m,n}+\dot\sigma_{bb}^{m,n}
+\dot\sigma_{cc}^{m,n})=\Gamma_R\sigma_{cc}\\
\label{bb5a}
I_d&=&\sum_{m,n}n(\dot\sigma_{aa}^{m,n}+\dot\sigma_{bb}^{m,n}
+\dot\sigma_{cc}^{m,n})=D-(D-D')\sigma_{bb}
\label{bb5b}
\end{eqnarray}
\end{mathletters}
where $\sigma_{ij}=\sum_{m,n}\sigma_{ij}^{m,n}$. It follows from 
Eq.~({\ref{bb5b}) that the variation of the detector current 
$\Delta I_d=I_d^{(0)}-I_d$ is given by Eq.~(\ref{a8}), where 
$\bar\sigma =\sigma_{bb}$. Thus, 
the point-contact measures directly the occupation of the left dot.   

Performing summation in Eqs.~(\ref{b5}) 
over the number of electrons arrived to the 
collectors ($m,n$), we obtain the following Bloch equations for the   
reduced density-matrix of the double-dot system:
\begin{mathletters}
\label{b6}
\begin{eqnarray}
&&\dot\sigma_{aa}  = -\Gamma_L\sigma_{aa}+\Gamma_R\sigma_{cc}
\label{b6a}\\
&&\dot\sigma_{bb}  = \Gamma_L\sigma_{aa}+i\Omega_0 (\sigma_{bc}-\sigma_{cb})
\label{b6b}\\
&&\dot\sigma_{cc}  = -\Gamma_R\sigma_{cc}-i\Omega_0 (\sigma_{bc}-\sigma_{cb})
\label{b6c}\\
&&\dot\sigma_{bc}  = i\epsilon\sigma_{bc}+i\Omega_0 (\sigma_{bb}-\sigma_{cc})
-\frac{1}{2}(\Gamma_R+\Gamma_d)\sigma_{bc},
\label{b6d}
\end{eqnarray}
\end{mathletters}
where $\Gamma_d$ is the dephasing rate generated by the detector,
Eq.~(\ref{c5}). 
These equations can be compared with those 
describing electron transport through the same system, but 
without detector\cite{gp,g,naz}. We find that the 
difference appears only in the nondiagonal 
density-matrix elements, Eq.~(\ref{b6d}). 
The latter includes an additional 
dissipation rate $\Gamma_d$ generated by the detector.

Solving Eqs.~(\ref{b6}) in the limit $t\to\infty$ we find 
the following expression for the current $I_s$, Eq.~(\ref{bb5a}), 
flowing through the double-dot system 
\begin{equation}
I_s=\frac{(\Gamma_R+\Gamma_d)\Omega_0^2}
{\displaystyle\epsilon^2+\frac{(\Gamma_R+\Gamma_d)^2}{4}
+\Omega_0^2(\Gamma_R+\Gamma_d)
\left (\frac{2}{\Gamma_R}+\frac{1}{\Gamma_L}\right )}
\label{b8}
\end{equation}

By analyzing Eq.~(\ref{b8}) one finds  
that all the measurement effects, discussed in the 
previous section are reflected in the behavior of the resonant current, 
$I_s$ as a function of the level displacement $\epsilon$ and the 
dephasing rate $\Gamma_d$. As an example, we show in Fig. 5 the 
resonant current $I_s(\epsilon )$ for three values of
the dephasing rate: $\Gamma_d=0$,  $\Gamma_d=4\Omega_0$ and 
$\Gamma_d=16\Omega_0$. We find that  
for small $\epsilon$ the current decreases with $\Gamma_d$,
while for large $t$ the average distribution of an electron 
in the dots remains the same. 
However, for larger values of $\epsilon$ the current 
{\em increases} with $\Gamma_d$.
It reflects electron delocalization in a double-well 
system, Fig. 3b, due to continuous monitoring of the 
charge in the left dot. In contrast, rapidly repeated 
measurement\cite{beige,fre} would always localize an electron and therefore 
diminish the current $I_s$. 
\vskip1cm
\begin{minipage}{13cm}
\begin{center}
\leavevmode
\epsfxsize=10cm
\epsffile{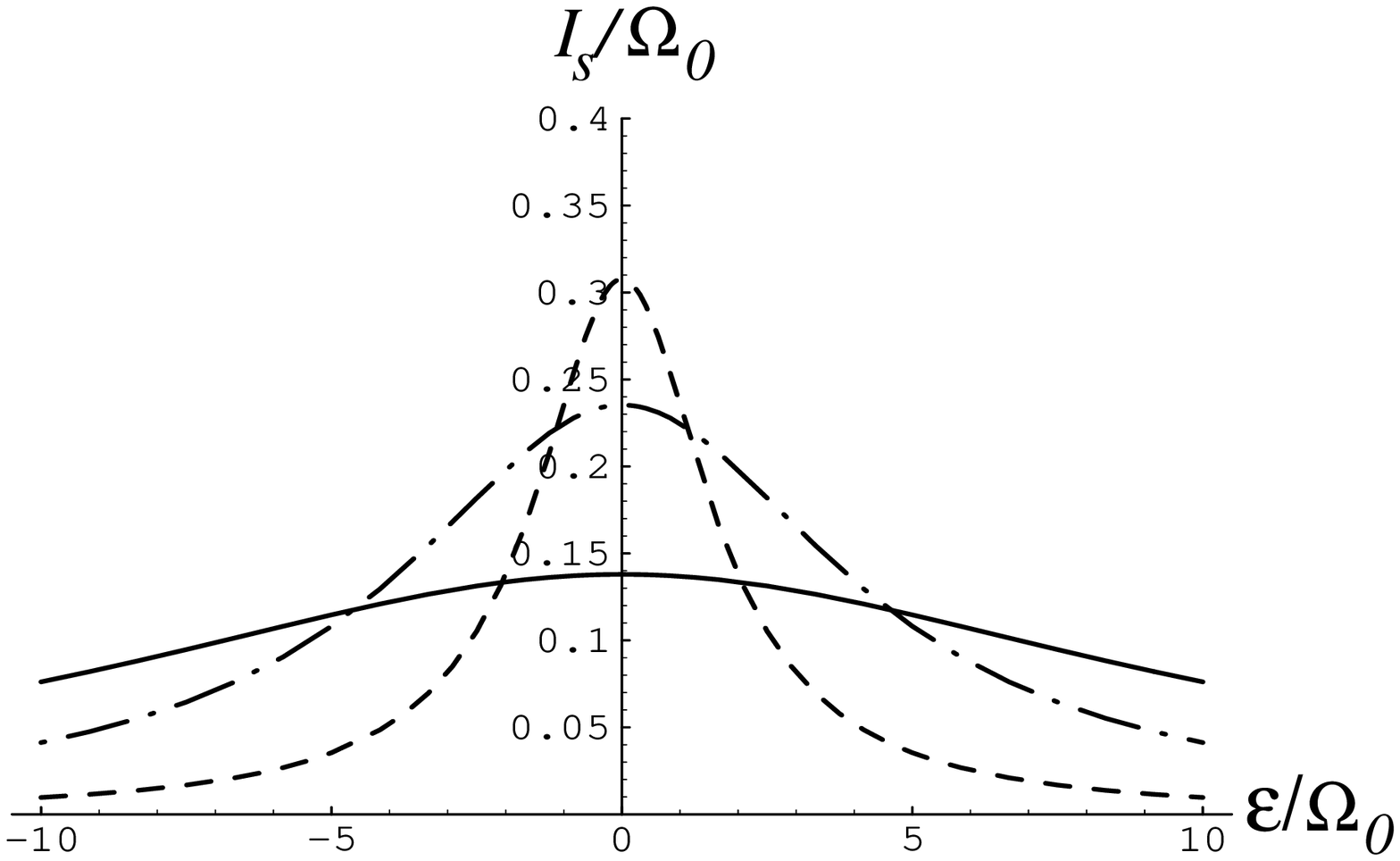}
\end{center}
{\begin{small}
Fig.~5. Electron current through the doubled-dot, Eq.~(\ref{b8}), 
for $\Gamma_L=\Gamma_R=\Omega_0$ 
as a function of the level displacement $\epsilon =E_2-E_1$.
The curves correspond to different values of the dephasing 
rate: $\Gamma_d=0$ (dashed),  $\Gamma_d=4\Omega_0$ (dot-dashed) and 
$\Gamma_d=16\Omega_0$ (solid). 
\end{small}}
\end{minipage} \\ \\ 

\section{Summary}
In this paper we studied the mechanism of decoherence 
generated by continuous observation of one of the states out of 
the coherent superposition in experiments with mesoscopic systems. 
As an example we considered a coupled quantum-dot system, 
which is simple enough 
for detailed theoretical treatment of the measured object 
and the detector together. 
On the other hand, it bears all essential physics of the measurement 
process. For a description of the entire system we applied 
the Bloch-type equations, which are obtained from the 
many-body Schr\"odinger equation and provide the most 
simple and transparent treatment of quantum coherence effects.
 
As  the detector we used the point contact in close
proximity to one of the dots. We demonstrated that 
the variation of the point-contact current 
due to electrostatic interaction with electrons in the dot 
measures directly the occupation of this dot. 

We started with quantum  
oscillations of an electron in coupled quantum dots. 
It appears that the presence of the point-contact 
detector near one of the dots generates the dephasing 
rate in the Bloch equations for  the off-diagonal density matrix 
elements. We found that the dephasing rate is
proportional to the variation of the point-contact transmission 
amplitude squared, Eq~(\ref{c5}).
The Bloch equations for the diagonal density-matrix elements 
are not affected by the detector, providing that  
it does not distort the energy levels of 
the double-dot system. 

The appearance of the dephasing rate $\Gamma_d$ in the 
Bloch equation leads to the collapse of the density 
matrix into the statistical mixture at $t\to\infty$, Eq.~(\ref{c6}).
The collapse happens even for large disalignment of the energy 
levels. In this case the measurement process results in 
an electron delocalization inside the double-dot (after 
some critical time $t>t_0$), which otherwise would stay localized in 
one of the dots. It contradicts to a common opinion that the continuous 
measurement always leads to a localization due to the wave-packet 
reduction (Zeno effect). In fact the localization would take 
place if we consider the continuous measurement as rapidly repeated 
measurements with intervals $\Delta t=t/n$ for $n\to\infty$. 
The reason of such a different behavior of the measured system 
stems from the different procedure of tracing out of the detector 
variables from the total density matrix.  

The same measurement effects appear in the dc current flowing 
through coupled-dots. We found that  
the dc current vanishes for $\Gamma_d\to\infty$, 
which can be interpreted in terms of an electron 
localization due to the Zeno effect. 
Nevertheless, for a finite $\Gamma_d$ 
and for a disaligned energy levels ($E_1\not =E_2$) the dc current  
{\em increases} with $\Gamma_d$. Here again,  
the situation is opposite to that of rapidly repeated measurement, where 
the current always {\em decreases} with $\Gamma_d$.

\section{Acknowledgments}
I owe special thanks to E. Buks for fruitful discussions on various 
theoretical and experimental aspects. I am also grateful to M. Heiblum and  
Y. Levinson for useful discussions. 

\appendix 
\section{rate equations for a point-contact detector}
We present here the microscopic derivation of the rate equations describing 
electron transport in the point contact. 
The point-contact is considered as a barrier, separated two reservoirs
(the emitter and the collector), Fig. 1.
All the levels in the emitter and the collector 
are initially filled up to the Fermi 
energies $\mu_L$ and $\mu_R$ respectively. We call 
it as the ``vacuum'' state, $|0\rangle$.
The tunneling Hamiltonian ${\cal H}_{pc}$ describing this system 
can be written as 
\begin{equation}
{\cal H}_{PC}=\sum_l E_la_l^\dagger a_l+\sum_r E_ra_r^\dagger a_r
+\sum_{l,r}\Omega_{lr}(a_l^\dagger a_r+H.c.), 
\label{ap1}
\end{equation}
where $a_l^\dagger (a_l)$ and $a_r^\dagger  (a_r)$ are the creation
(annihilation) operators in the left and the right 
reservoirs, respectively, and $\Omega_{lr}$ is the hopping amplitude 
between the states $E_l$ and $E_r$ in the right and the left reservoirs.  
(We choose the the gauge where $\Omega_{lr}$ is real).
The Hamiltonian Eq.~(\ref{ap1}) requires the vacuum state $|0\rangle$ to decay 
exponentially to a continuum states 
having the form: $a_{r}^{\dagger}a_{l}|0\rangle$ with an electron
in the collector continuum 
and a hole in the emitter continuum; 
$a_{r}^{\dagger}a_{r'}^{\dagger}a_{l}^{\dagger}a_{l'}|0\rangle$ 
with two electrons in the collector continuum and two holes in 
the emitter continuum, and so on. 
The many-body wave function describing this system can be written 
in the occupation number representation as 
\begin{equation}
|\Psi (t)\rangle = \left [ b_0(t) + \sum_{l,r} b_{lr}(t)a_r^{\dagger}a_l
+\sum_{l<l',r<r'} b_{ll'rr'}(t)a_r^{\dagger}a_{r'}^{\dagger}a_la_{l'}
+\cdots\right ]|0\rangle, 
\label{ap2}
\end{equation}
where $b(t)$ are the time-dependent probability amplitudes to
find the system in the corresponding states with the initial 
condition $b_0(0)=1$, and all the other $b(0)$'s being zeros.
Substituting Eq.~(\ref{ap2}) into the Shr\"odinger equation 
$i|\dot\Psi (t)\rangle ={\cal H}_{PC}|\Psi (t)\rangle$ 
and performing the Laplace transform, 
\begin{equation}
\tilde{b}(E)=\int_0^{\infty}e^{iEt}b(t)dt
\label{ap3}
\end{equation}
we obtain an infinite set of the coupled equations for the 
amplitudes $\tilde b(E)$:
\begin{mathletters}
\label{ap4}
\begin{eqnarray}
& &E \tilde{b}_{0}(E) - \sum_{l,r} \Omega_{lr}\tilde{b}_{lr}(E)=i
\label{ap4a}\\
&(&E + E_{l} - E_r) \tilde{b}_{lr}(E) - \Omega_{lr}\tilde{b}_0(E) 
-\sum_{l',r'}\Omega_{l'r'}\tilde{b}_{ll'rr'}(E)=0
\label{ap4b}\\
&(&E + E_{l}+E_{l'} - E_r-E_{r'}) \tilde{b}_{ll'rr'}(E) 
- \Omega_{l'r'}\tilde{b}_{lr}(E)+\Omega_{lr}\tilde{b}_{l'r'}(E)
\nonumber\\
&&~~~~~~~~~~~~~~~~~~~~~~~~~~~~~~~~~~~~~~~~~~~~~~~~~~~~~
-\sum_{l'',r''}\Omega_{l''r''}\tilde{b}_{ll'l''rr'r''}(E)=0
\label{ap4c}\\
& &\cdots\cdots\cdots\cdots\cdots 
\nonumber
\end{eqnarray}
\end{mathletters}
Eqs. (\ref{ap4}) can be substantially simplified by replacing 
the amplitude $\tilde b$ in 
the term $\sum\Omega\tilde b$ of each of the equations  by 
its expression obtained from the subsequent equation\cite{gp,g}.  
For example,   
substituting $\tilde{b}_{lr}(E)$ from Eq.~(\ref{ap4b}) into Eq.~(\ref{ap4a}), 
one obtains
\begin{equation}
\left [ E - \sum_{l,r}\frac{\Omega^2}{E + E_{l} - E_r}
    \right ] \tilde{b}_{0}(E) - \sum_{ll',rr'}
    \frac{\Omega^2}{E + E_{l} - E_r}\tilde{b}_{ll'rr'}(E)=i,
\label{ap5}
\end{equation}
where we assumed that the hopping amplitudes 
are weakly dependent functions on the energies
$\Omega_{lr}\equiv\Omega (E_l,E_r)=\Omega$.
Since the states in the reservoirs are very dense (continuum), 
one can replace the sums over $l$ and $r$ by integrals, for instance  
$\sum_{l,r}\;\rightarrow\;\int \rho_{L}(E_{l})\rho_{R}(E_{r})\,dE_{l}dE_r\:$,
where $\rho_{L,R}$ are the density of states in the emitter and collector. 
Then the first sum in Eq.~(\ref{ap5}) becomes an
integral which can be split into a sum of the singular and principal value 
parts. The singular part yields $i\pi\Omega^2\rho_L\rho_R V_d$,
and the principal part is merely included into
redefinition of the energy levels. The second sum in Eq.~({\ref{ap5}) 
can be neglected. Indeed, by replacing 
$\tilde{b}_{ll'rr'}(E)\equiv \tilde{b} (E,E_l,E_{l'},E_r,E_{r'})$ and 
the sums by the integrals we find that the integrand  
has the poles on the same sides of the integration 
contours. It means that the corresponding integral vanishes, providing 
$V_d\gg \Omega^2\rho$.  

Applying analogous considerations to the other equations of the
system (\ref{ap4}), we finally arrive to the following set of equations: 
\begin{mathletters}
\label{ap6}
\begin{eqnarray}
&& (E + iD/2) \tilde{b}_{0}=i
\label{a6a}\\
&& (E + E_{l} - E_r + iD/2) \tilde{b}_{lr}
      - \Omega\tilde{b}_{0}=0
\label{ap6b}\\ 
&& (E + E_{l}+ E_{l'} - E_{r} - E_{r'} + iD/2) \tilde{b}_{ll'rr'} -
      \Omega\tilde{b}_{lr}+\Omega \tilde{b}_{l'r'}=0,
\label{a6c}\\
& &\cdots\cdots\cdots\cdots\cdots 
\nonumber
\end{eqnarray}
\end{mathletters}
where $D=2\pi\Omega^2\rho_L\rho_R V_d$. 

The charge accumulated in the collector at time $t$ is
\begin{equation}
N_R(t) =\langle\Psi (t)|\sum_r a_r^\dagger a_r|\Psi (t)\rangle=
\sum_nn\sigma^{(n)}(t),
\label{ap7}
\end{equation} 
where 
\begin{equation}
\sigma^{(0)}(t)=|b_0(t)|^2,~~~~\sigma^{(1)}(t)=\sum_{l,r}|b_{lr}(t)|^2,~~~~
\sigma^{(2)}(t)=\sum_{ll',rr'}|b_{ll'rr'}(t)|^2,\; \cdots
\label{ap8}
\end{equation}  
are the probabilities to find $n$ electrons in the collector. 
These probabilities are directly related 
to the amplitudes $\tilde b(E)$ through the inverse Laplace transform 
\begin{equation}
\sigma^{(n)}(t)=
\sum_{l\ldots , r\ldots}
\int \frac{dEdE'}{4\pi^2}\tilde b_{l\cdots r\cdots}(E)
\tilde b^*_{l\cdots r\cdots}(E')e^{i(E'-E)t}
\label{ap9}
\end{equation}
Using Eq.~(\ref{ap9}) one can transform  Eqs.~(\ref{ap6}) into the rate
equations for $\sigma^{(n)}(t)$ (c.f.\cite{gp,g}). We find 
\begin{mathletters}
\label{ap10}
\begin{eqnarray}  
&&\dot{\sigma}^{(0)}(t) = -D\sigma^{(0)}(t)
\label{a10a}\\
&&\dot{\sigma}^{(1)}(t) = D\sigma^{(0)}(t)-D\sigma^{(1)}(t)
\label{a10b}\\
&&\dot{\sigma}^{(2)}(t) = D\sigma^{(1)}(t)-D\sigma^{(2)}(t)
\label{a10c}\\
&&\cdots\cdots\cdots\cdots\cdots 
\nonumber
\end{eqnarray}
\end{mathletters}

The operator, which defines the current flowing in this system is 
\begin{equation}
\hat I=i\left [{\cal H}_{PC},\sum_r a_r^\dagger a_r\right ]=
i\sum_{l,r}\Omega_{lr}(a^\dagger_la_r
-a^\dagger_ra_l)
\label{ap11}
\end{equation}
Using Eqs.~(\ref{ap2}), (\ref{ap10}) and  (\ref{ap11}) we find for 
the current
\begin{equation} 
I =\langle\Psi (t)|\hat I|\Psi (t)\rangle=
D\sum_n\sigma^{(n)}(t)=D.
\label{ap12}
\end{equation}
Since $D=(2\pi )^2\Omega^2\rho_L\rho_R=T$\cite{bardeen}, 
where $T$ is the transmission probability, 
the current can be rewritten as $I=T\, V_d/(2\pi )$, which is 
the well known Landauer formula. 

\section{Point-contact detector near double-well}
Now we present the microscopic derivation of the Bloch equations 
(\ref{c3}) describing electron oscillations in a double-well with 
a point-contact in close proximity to one of the wells, Fig. 2.  
We start with the many-body Schr\"odinger equation 
$i|\dot\Psi (t)\rangle ={\cal H}|\Psi (t)\rangle$ for 
the entire system. Here ${\cal H}$ is 
the tunneling Hamiltonian, which can be written as 
${\cal H}={\cal H}_{PC}+{\cal H}_{DD}+{\cal H}_{int}$.
Here ${\cal H}_{PC}$ is the tunneling Hamiltonian for 
the point-contact detector, 
Eq.~(\ref{ap1}); ${\cal H}_{DD}$ is tunneling Hamiltonian for 
the measured double-dot system,
\begin{equation}
{\cal H}_{DD} = E_1 c_1^{\dagger}c_{1}+E_2 c_2^{\dagger}c_{2}+
              \Omega_0 (c_2^{\dagger}c_{1}+ c_1^{\dagger}c_{2})\, ,
\label{bp1}
\end{equation}
and ${\cal H}_{int}$ describes the interaction between the 
detector and the measured system. Since the presence of an electron in the 
left well results in an effective increase of the point-contact barrier 
($\Omega_{lr}\to\Omega_{lr}+\delta\Omega_{lr}$), we can 
represent the interaction term as 
\begin{equation}
{\cal H}_{int}=\sum_{l,r}\delta\Omega_{lr}c_1^{\dagger} 
c_1(a^{\dagger}_la_r+H.c.).
\label{bp2}
\end{equation}  
The many-body wave function for the entire system can be written as
\begin{eqnarray}
|\Psi (t)\rangle &=& \left [ b_1(t)c_1^{\dagger} 
+ \sum_{l,r} b_{1lr}(t)c_1^{\dagger}a_r^{\dagger}a_l
+\sum_{l<l',r<r'} b_{1ll'rr'}(t)
c_1^{\dagger}a_r^{\dagger}a_{r'}^{\dagger}a_la_{l'}\right.
\nonumber\\
&+&\left. b_2(t)c_2^{\dagger}
+ \sum_{l,r} b_{2lr}(t)c_2^{\dagger}a_r^{\dagger}a_l
+\sum_{l<l',r<r'} b_{2ll'rr'}(t)
c_2^{\dagger}a_r^{\dagger}a_{r'}^{\dagger}a_la_{l'}+\cdots
\right ]|0\rangle,
\label{bp3}
\end{eqnarray} 
where $b(t)$ are the probability amplitudes to find the entire 
system in the states defined by the corresponding creation and 
annihilation operators. Notice that Eq.~(\ref{bp3}) has the same
form as Eq.~(\ref{ap2}), where only the probability amplitudes $b(t)$ 
acquire an additional index ('1' or '2') that denotes the well, occupied 
by an electron. Proceeding in the same way as in Appendix A
we arrive to an infinite set of the coupled equations for the 
amplitudes $\tilde b(E)$, which are the Laplace transform 
of the amplitudes $b(t)$, Eq.~(\ref{ap3}):
\begin{mathletters}
\label{bp4}
\begin{eqnarray}
& &(E-E_1) \tilde{b}_{1}(E) - \Omega_0\tilde b_2(E) 
-\sum_{l,r} \Omega'_{lr}\tilde{b}_{1lr}(E)=i
\label{bp4a}\\
& &(E-E_2) \tilde{b}_{2}(E)  - \Omega_0\tilde b_1(E)
- \sum_{l,r} \Omega_{lr}\tilde{b}_{2lr}(E)=0
\label{bp4b}\\
&(&E + E_{l}-E_1 - E_r) \tilde{b}_{1lr}(E) - \Omega'_{lr}\tilde{b}_1(E)
-\Omega_0\tilde b_{2lr}(E) 
-\sum_{l',r'}\Omega_{l'r'}\tilde{b}_{1ll'rr'}(E)=0
\label{bp4c}\\
&(&E + E_{l}-E_2 - E_r) \tilde{b}_{2lr}(E) - \Omega_{lr}\tilde{b}_2(E)
-\Omega_0\tilde b_{1lr}(E) 
-\sum_{l',r'}\Omega_{l'r'}\tilde{b}_{2ll'rr'}(E)=0
\label{bp4d}\\
& &\cdots\cdots\cdots\cdots\cdots 
\nonumber
\end{eqnarray}
\end{mathletters}
The same algebra as that used in the Appendix A and in 
Refs.\cite{gp,g} allows us to 
simplify these equations, which then become 
\begin{mathletters}
\label{bp5}
\begin{eqnarray}
&& (E -E_1+ iD'/2) \tilde{b}_{1}-\Omega_0\tilde b_2=i
\label{bp5a}\\
&& (E -E_2+ iD/2) \tilde{b}_{2}-\Omega_0\tilde b_1=0
\label{bp5b}\\
&& (E + E_{l} -E_1- E_r + iD'/2) \tilde{b}_{1lr}
      - \Omega'\tilde{b}_{1}-\Omega_0\tilde b_{2lr}=0
\label{bp5c}\\ 
&& (E + E_{l} -E_2- E_r + iD/2) \tilde{b}_{2lr}
      - \Omega\tilde{b}_{2}-\Omega_0\tilde b_{1lr}=0
\label{bp5d}\\ 
& &\cdots\cdots\cdots\cdots\cdots 
\nonumber
\end{eqnarray}
\end{mathletters}
where $D=TV_d/2\pi$. 
(We assumed for simplicity that the hopping amplitude of the point-contact is  
weakly dependent on the energies, so that 
$\Omega_{lr}\equiv\Omega (E_l,E_r)=\Omega$).

Using the inverse Laplace transform (\ref{ap9}) we can transform 
Eqs.~(\ref{bp5}) into differential equations for the density-matrix 
elements $\sigma^{(n)}_{ij}(t)$ ($i,j$=1,2)
\begin{equation}
\sigma^{(0)}_{ij}(t)=b_i(t)b^*_j(t),
~~~\sigma^{(1)}_{ij}(t)=\sum_{l,r}b_{ilr}(t)b^*_{jlr}(t),
~~~\sigma^{(2)}_{ij}(t)=
\sum_{ll',rr'}b_{ill'rr'}(t)b^*_{jll'rr'}(t),\; \cdots\ ,
\label{bp6}
\end{equation}  
where $n$ denotes the number of electrons 
accumulated in the collector. Consider, for instance the off-diagonal 
density-matrix element $\sigma^{(1)}_{12}(t)$. The corresponding 
differential equation for this term can by obtained by 
multiplying Eq.~(\ref{bp5c}) by 
$\tilde b^*_{2lr}(E')$ and subtracting the complex conjugated 
Eq.~(\ref{bp5d}) multiplied by $\tilde b_{1lr}(E)$. We then 
obtain
\begin{eqnarray} 
&&\int\frac{dEdE'}{4\pi^2}\sum_{l,r}\left\{\left (E'-E-\epsilon 
-i\frac{D+D'}{2}\right )\tilde b_{1lr}(E)\tilde b^*_{2lr}(E')
\right.\nonumber\\
&&\left.~~~~~~~~~~~~~~~~~~~~~~~~~~~~
-[\Omega \tilde b_{1lr}(E)\tilde b^*_2(E')
-\Omega' \tilde b^*_{2lr}(E')\tilde b_1(E)]
\right.\nonumber\\
&&\left.~~~~~~~~~~~~~~~~~~~~~~~~~~~~~~~~~~
-\Omega_0[\tilde b_{1lr}(E)\tilde b^*_{1lr}(E')
-\tilde b^*_{2lr}(E')\tilde b_{2lr}(E)]
\right\}e^{i(E'-E)t}=0.
\label{bp7}
\end{eqnarray}
One easily finds that the first term 
in this equation equals to  
$-i\dot\sigma_{12}^{(1)}-[\epsilon+i(D+D')/2]\sigma_{12}^{(1)}$ and 
the third term equals to $-\Omega_0(\sigma_{11}^{(1)}-
\sigma_{22}^{(1)})$. In order to evaluate the second term in Eq.~(\ref{bp7})
we replace $\sum_{l,r}$ by the integrals and substitute  
\begin{eqnarray}
\tilde b_{1lr}(E)=\frac{\Omega'\tilde b_1(E)+\Omega_0\tilde b_{2lr}(E)}
{E+E_l-E_1-E_r+iD'/2}\nonumber\\
\tilde b^*_{2lr}(E')=\frac{\Omega\tilde b^*_2(E')
+\Omega_0\tilde b^*_{1lr}(E')}
{E'+E_l-E_2-E_r-iD/2}
\end{eqnarray}
obtained from Eqs.~(\ref{bp5c}), (\ref{bp5d}), 
into Eq.~(\ref{bp7}). Then integrating over $E_l,E_r$ we find that 
the second term in Eq.~(\ref{bp7}) becomes 
$2i\pi\Omega\Omega'\rho_L\rho_RV_d\sigma_{12}^{(0)}$. Thus  
Eq.~(\ref{bp7}) can be rewritten as 
\begin{equation}  
\dot\sigma_{12}^{(1)} = i\epsilon\sigma_{12}^{(1)}+
i\Omega_0(\sigma_{11}^{(1)}-\sigma_{22}^{(1)})
-\frac{1}{2}(D'+D)\sigma_{12}^{(1)}
+(D\, D')^{1/2}\sigma_{12}^{(0)}\;.
\label{bp9}
\end{equation}
which coincides with the Bloch equation (\ref{c3c}) for $n=1$ and 
$\sigma_{aa}\equiv\sigma_{11}$, $\sigma_{bb}\equiv\sigma_{22}$,
$\sigma_{ab}\equiv\sigma_{12}$. Applying the same procedure 
to each of the equations (\ref{bp5}) we arrive to the 
Bloch equations (\ref{c3}) for density matrix elements $\sigma_{ij}^{(n)}$.


\begin{references}
\bibitem{fn1} Note that  $\Gamma_d\simeq (\Delta T)^2V_d/8\pi T$ for
$\Delta T\ll T$, where $\Delta T=T-T'$. It is different from a similar 
expression for the dephasing rate found in\cite{awm,levin} 
by a factor $(1-T)$.
\bibitem{fn3} Notice that the previous investigations 
of the Zeno effect in atomic system\cite{beige,fre}
always considered the situation corresponding to alignment 
leves ($E_1=E_2$) in our case. 
As a result the average probability of finding an electron 
on one of the levels remains the same before and after  
the density-matrix collapse.
\bibitem{fn4} Notice, that our Bloch-type equations (\ref{c3}),
(\ref{c4}) cannot be mapped to the Bloch equations 
for 3-level system\cite{fre} due to continuum-to-continuum 
transitions in the point-contact detector.
\end{references}
\end{document}